\documentclass[runningheads]{llncs}
\usepackage{graphicx}
\usepackage{amsmath,amsthm,amssymb}

\usepackage{xcolor}

\usepackage{subfig}

 \usepackage{multirow}
 \usepackage{colortbl}
  \usepackage{rotating}
  
\usepackage[hidelinks]{hyperref}

\newcommand\blfootnote[1]{%
  \begingroup
  \renewcommand\thefootnote{}\footnote{#1}%
  \addtocounter{footnote}{-1}%
  \endgroup
}

\begin{document}
\title{Diabetes Link: Platform for Self-Control and Monitoring People with Diabetes}
\titlerunning{Diabetes Link}
%
\author{Enzo Rucci\inst{1,3}\orcidID{0000-0001-6736-7358}\thanks{Corresponding author.} \and
Lisandro Delía\inst{1}\orcidID{0000-0003-0515-0609} \and
Joaquín Pujol\inst{2}\and
Paula Erbino\inst{2}\and
Armando De Giusti\inst{1}\orcidID{0000-0002-6459-3592}\and
Juan José Gagliardino\inst{3}}
%
%
\institute{III-LIDI, Facultad de Informática, UNLP – CeAs CICPBA. \\La Plata (1900), Bs As, Argentina \\
\email{erucci@lidi.info.unlp.edu.ar} \\
 \and
Facultad de Informática, UNLP. \\
La Plata (1900), Bs As, Argentina\\
\and
CENEXA, UNLP – CONICET – CeAs CICPBA. \\La Plata (1900), Bs As, Argentina \\
}

\maketitle              
\begin{abstract}

Diabetes Mellitus (DM) is a chronic disease characterized by an increase in blood glucose (sugar) above normal levels and it appears when human body is not able to produce enough insulin to cover the peripheral tissue demand. Nowadays, DM affects the 8.5\% of the world’s population and, even though no cure for it has been found, an adequate monitoring and treatment allow patients to have an almost normal life.    This paper introduces Diabetes Link, a comprehensive platform for control and monitoring people with DM. Diabetes Link allows recording various parameters relevant for the treatment and calculating different statistical charts using them. In addition, it allows connecting with other users (supervisors) so they can monitor the controls. Even more, the extensive comparative study carried out reflects  that Diabetes Link presents distinctive and superior features against other proposals. We conclude that Diabetes Link represents a broad and accessible tool that can help make day-to-day control easier and optimize the efficacy in DM control and treatment.~\blfootnote{\texttt{The final authenticated version is available online at \url{https://doi.org/10.1007/978-3-030-61702-8\_25}}}

\keywords{Diabetes \and Diabetes control \and Blood sugar monitoring \and Health informatics \and Mobile application  \and eHealth   \and mHealth }
\end{abstract}

\section{Introduction}
\label{sec:intro}

According to the American Institute of Medical Sciences \& Education, the term mobile health (\textit{mHealth}) is used to refer to mobile technology focused on health care and medical information. There are currently more than 100,000 mobile applications associated with the health field. In the United States, 80\% of medical professionals use mobile phones and medical applications, 25\% of which are used for patient care~\cite{TheImpactOfTechnologyInHealthcare}.

Diabetes Mellitus (DM) is a chronic disease characterized by an increase in blood glucose (sugar) above normal levels~\cite{LibroComoTratarMiDiabetes}. Affected people need ongoing and long-term treatment, as well as periodic self-monitoring of blood glucose (SMBG) mainly, but also of other clinical and metabolic parameters, such as body weight, food intake (measured mainly in carbohydrates), blood pressure, medications used, insulin dose and physical activity performed. Failure to adequately control the disease or comply with the prescribed treatments can not only have irreparable consequences on overall health, but it can also affect the quality of life of people with diabetes (namely, chronic complications affecting the eyes, kidneys and heart, among other organs; as well as psychological and mental damage). At the same time, it also affects patients from and economic and social perspective, due to the high costs of treatment and the impact on work discrimination~\cite{Elgart2014}.

Globally, according to the latest figures presented by the World Health Organization (WHO) in October 2018, the number of people with diabetes has increased significantly~\cite{DiabetesOMS}. In 1980, it affected 4.7\% of the world's population, and this increased to 8.5\% by 2014, with a total of 422 million people affected. Among people with diabetes, those with type 1 diabetes (T1D) and some with type 2 diabetes (T2D) receive daily insulin injections and perform SMBG to define the insulin dose they need to inject and optimize blood sugar control. This process is required to reduce the development and progression of the chronic complications of the disease  and its negative impact in quality of life.

In this paper, we present \textit{Diabetes Link}, a comprehensive platform to simplify the control and monitoring of people with DM. Diabetes Link combines a mobile application and a web portal, making the best of each platform. It allows recording various parameters relevant for the appropiate treatment and calculating different statistical charts using them. Also, it allows connecting with other users (supervisors) so they can monitor the controls. In addition, Diabetes Link offers all its functionality for free, and it is available on multiple platforms. Through an extensive comparative analysis, we  show the distinctive and superior features of Diabetes Link against other available tools.

The remaining sections of this article are organized as follows: Section 2 describes the recommended treatment for people with DM. Section 3 introduces Diabetes Link, and Section 4 presents a comparative analysis with other available options. Finally,  Section 5 summarizes the conclusions and and possible future lines of work.

\section{Treatment for Diabetes Mellitus}
\label{sec:dm}

So far, no cure for DM has been found, but sustained control and adequate treatment allow patients to lead an almost normal life. In addition, an active participation of the patients in the control and treatment of their disease is required to achieve this goal.  

Treatment is based on four pillars – education, healthy meal plan, regular practice of physical activity, and various medications (oral antidiabetic drugs and insulin if required). It is very important that people with diabetes keep their condition under continuous and tight control and adhere to their prescribed treatment throughout their lives. In addition, depending on patient and disease severity, different clinical and metabolic parameters should be monitored, such as body weight, blood glucose, blood pressure, the amount of ingested carbohydrates, and physical activity performed. All these parameters have an impact on the long-term level of control and the simultaneous monitoring prevents the development/progression of the chronic complications that deteriorate quality of life and increase their costs of care~\cite{Elgart2014,LibroComoTratarMiDiabetes,Steno2}.

Recording the various parameters associated with the treatment allows physicians to monitor patient evolution over time and to adjust the treatment accordingly in order to optimize results. In that sense, the ultimate goal is that the patient can maintain certain degree of stability. 

The lack of an adequate control of the disease, together with non-adherence to prescribed treatments, leads to the development and progression of the chronic complications of the disease. They affect various organs such as the retina, kidneys, and cardiovascular system, decreasing their quality of life and their psychophysical ability as well as increasing their costs of care. Equally important, the decrease in psychophysical capacity also affects the ability to work, with the consequent socio-economic impact.

\section{Diabetes Link}
\label{sec:dlink}

\subsection{Work Methodology}


To analyze the potential of the proposal and subsequently define its scope, clinical researchers from the CENEXA (UNLP-CONICET-CeAs CICPBA)~\footnote{\url{www.cenexa.org}} were contacted, due to their long career devoted to diabetes epidemiology, care programmes and derived costs. Their advice allowed us to learn about basic concepts of the disease and multiple aspects of its treatment. In addition, they emphasized the importance of keeping an adequate record of relevant data to achieve a good metabolic control. 


Two physicians specialized in DM (with experience in therapeutic education of people with diabetes and co-authors of a manual oriented to the complementation of this activity~\cite{LibroComoTratarMiDiabetes}) were also interviewed. They provided useful information related to the statistical indicators that specialists frequently analyze when assessing the status of a patient and adjusting their treatment.


Finally, this development phase was complemented with a state-of-the-art study of mobile applications oriented to diabetes, which allowed us to discover possible uncovered needs in the area. It is important to remark that this study was repeated once again after the platform was released to carry out the comparative analysis of Section~\ref{sec:results}.

\subsection{Requirements Analysis and Design}


For the first release of Diabetes Link, a set of functional and non-functional requirements of the platform was agreed among all the participants. Requirements analysis and design modelling were required phases in the plataform development to guarantee the quality of the final product.

\subsubsection{Functional Requirements} 

Table~\ref{tab:func-reqs} shows the functional requirements  of Diabetes Link in a simplified manner.

\subsubsection{Use Cases Model}

\begin{figure}[h!]%
    \centering
    \includegraphics[width=.9\textwidth]{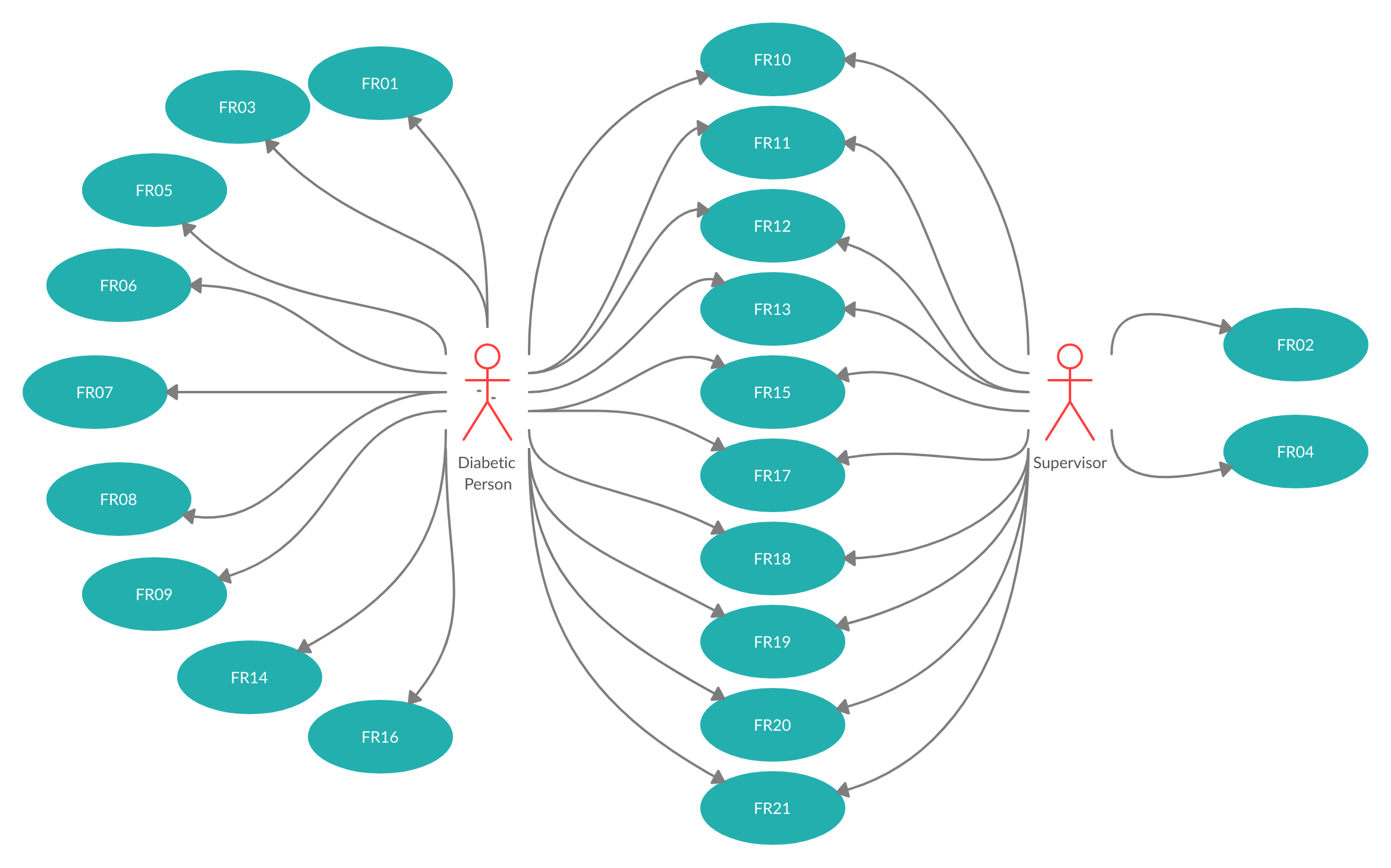} 
    \caption{Use cases}%
    \label{fig:usecases}%
\end{figure}

Fig.~\ref{fig:usecases} shows the use cases model of Diabetes Link. It is important to note that as these functional requirements are fine-grained, it was possible to translate each one into a particular use case.

\subsubsection{Non-Functional Requirements} 

Table~\ref{tab:non-func-reqs} shows the non-functional requirements  of Diabetes Link.

\begin{table}[h!]
\caption{\label{tab:func-reqs} Functional Requirements  of Diabetes Link}
\centering
\resizebox{\textwidth}{!}{%
\begin{tabular}{p{0.1\textwidth}p{0.7\textwidth}>{\centering\arraybackslash}p{0.1\textwidth}>{\centering\arraybackslash}p{0.1\textwidth}}
\hline
{ \textbf{ID}} &  \textbf{Functional Requirement}                                                                                                                                                                      & { \textbf{Web}} & { \textbf{Mobile}} \\
 &  \textit{The systems must allow the} & { \textbf{app}} & { \textbf{app}} \\                                
\hline
{ FR01}        & { Login/Logout of patient/supervised user}                                                                                                                                                                                               & { }             & { \checkmark}               \\
{ FR02}        & { Login/Logout of physician/supervisor user}                                                                                                                                                                                                         & { \checkmark}            & { \checkmark}               \\
{ FR03}        & { Creation of supervised user}                                                                                                                                                                                               & { }             & { \checkmark}               \\
{ FR04}        & { Creation of supervisor user}                                                                                                                                                                                                         & { }             & { \checkmark}               \\
{ FR05}        & { Record of relevant information for metabolic control: blood glucose measurements, carbohydrates intake and injected insulin. For each variable, the time of day (before/after breakfast/lunch/snack/dinner) must be included, and additional comments can be added to help understand the information entered } & { }             & { \checkmark}               \\
{ FR06}        & { Record of medications}                                                                                                                                                                                                           & { }             & { \checkmark}               \\
{ FR07}        & { Record of physical activity (detailing intensity and duration) }                                                                                                                                                        & { }             & { \checkmark}               \\
{ FR08}        & { Record of body weight}                                                                                                                                                                                                          & { }             & { \checkmark}               \\
{ FR09}        & { Record of blood pressure}                                                                                                                                                                                                       & { }             & { \checkmark}               \\
{ FR10}        & { Analysis of the evolution of blood glucose}                                                                                                                                                                                            & { \checkmark}            & { \checkmark}               \\
{ FR11}        & { Analysis of the evolution of body weight and BMI}                                                                                                                                                                                   & { \checkmark}            & { \checkmark}               \\
{ FR12}        & { Analysis of the evolution of blood pressure}                                                                                                                                                                                    & { \checkmark}            & { \checkmark}               \\
{ FR13}        & { Weekly summary of daily records detailing, for each day and meal, physical activity,  carbohydrates intake, and blood glucose and injected insulin measurements before and after each intake}        & { \checkmark}            & { \checkmark}               \\
{ FR14}        & { Supervisor search}                                                                                                                                                                                                           & { }             & { \checkmark}               \\
{ FR15}        & { Association/Dissociation of supervisors}                                                                                                                                                                                              & { }             & { \checkmark}               \\
{ FR16}        & { List of supervisor users}                                                                                                                                                                                                            & { \checkmark}            & { \checkmark}               \\
{ FR17}        & { Access to supervised user profile}                                                                                                                                                                                             & { \checkmark}            & { \checkmark}               \\
{ FR18}        & { Access to the FAQs section about DM. There, patients can read about the disease, including its causes, consequences, and specifications about treatment goals.}                                                                                                                       & { }             & { \checkmark}               \\
{ FR19}        & { Setting target values for blood glucose and blood pressure to customize the statistical analysis}                                                                                          & { \checkmark}             & { \checkmark}               \\
{ FR20}        & { Support for different units of measurement: User can choose their preferred units of measure for blood glucose (mg/dL or mmol/L) and weight (kg or lbs) }                                                                                                           & { }             & { \checkmark}               \\
{ FR21}        & { Support for different languages}                                                                                                                                                                                                & { \checkmark}             & { \checkmark}               \\ \hline
\end{tabular}%
}
\end{table}

\begin{table}[t!]
\caption{\label{tab:non-func-reqs} Non-functional Requirements of Diabetes Link}
\centering
\resizebox{\textwidth}{!}{%
\begin{tabular}{p{0.1\textwidth}p{0.9\textwidth}}
\hline
{ \textbf{ID}} & { \textbf{Non-functional Requirement}}  \\                                                              
\hline

{ NFR01}       & { Security. The system must ensure all information of the registered end users are secured and not accessible by other party.}                            \\
{ NFR02}       & { Security. The system should back up its data every 12 hours and the copies must be stored in a secure off-site location}                                                         \\
{ NFR03}       & { Usability. The system should provide a systematic, simple and user-friendly interfaces.}                                                                \\
{ NFR04}       & { Usability. The system should provide internationalization support (at least, English and Spanish).}                                                               \\
{ NFR05}       & { Platforms Constraint. The mobile application should work under Android and iOS  platforms. Also, a web application is needed for supervisors access.}   \\
{ NFR06}       & { The system must have a "Terms \& conditions" section.}                   \\                                                       
\hline
\end{tabular}%
}
\end{table}

\subsubsection{Architecture}

\begin{figure}[t!]%
    \centering
    \includegraphics[width=.75\textwidth]{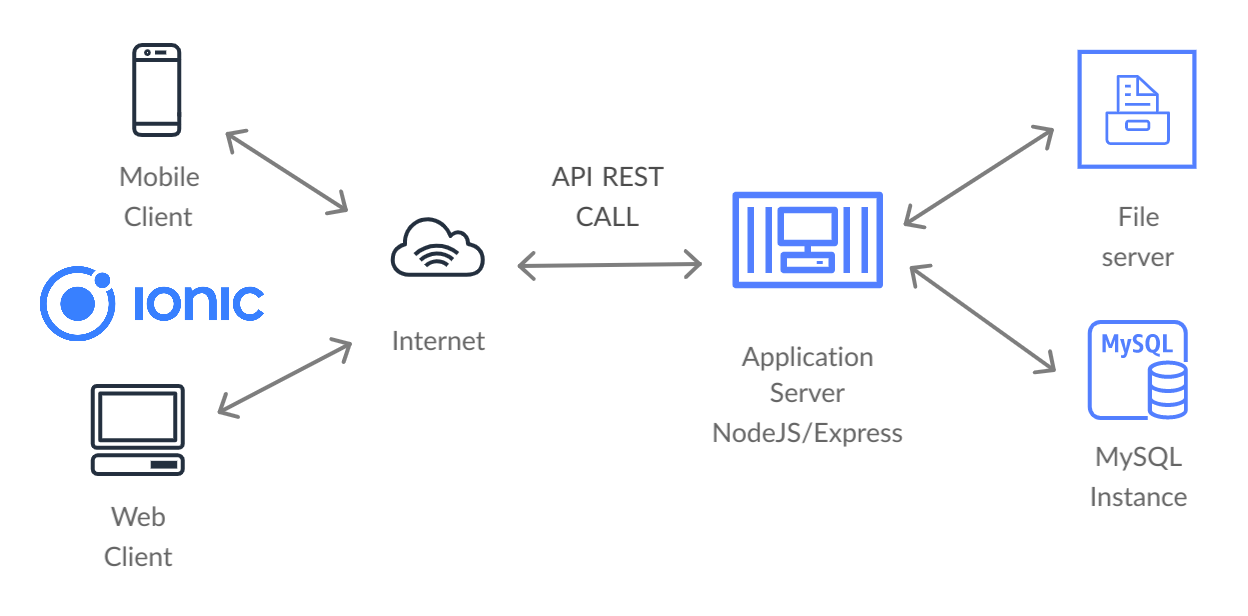} 
    \caption{General model of service-oriented architecture}%
    \label{fig:architecture}%
\end{figure}

The platform can be accessed through a mobile application (available now for Android~\footnote{\url{https://play.google.com/store/apps/details?id=lidi.diabetes.link}} and soon for iOS),  and/or a web application~\footnote{\url{http://diabeteslink.info.unlp.edu.ar}}. Figure~\ref{fig:architecture} shows a general model of service-oriented architecture.


The mobile application is used for both supervised and supervisor users. It was developed using web technologies (HTML5, CSS3 and Javascript) and the frameworks Angular 6~\footnote{\url{https://angularjs.org/}} and Ionic v3~\footnote{\url{https://ionicframework.com/docs/v3/}}. The latter, on its behalf, uses Apache Cordova~\footnote{\url{https://cordova.apache.org/docs/en/latest/guide/overview/index.html}} for packaging web solutions in modules that can be later installed in mobile devices. It is for that reason that the mobile application of Diabetes Link can be categorized as a multi-platform, hybrid one~\cite{Delia2015}. 

Nowadays, Android and iOS are the main mobile operating systems. The mobile application of Diabetes Link is ready to work on both of them due to its multi-platform, hybrid feature. In that sense, the Android version was released first allowing Diabetes Link to cover more than 70\% of the market share~\cite{MobileOSMarketShare}. On its behalf,  the iOS version will be released next.


Most of the mobile application code could be reused for the development of a web application, which is now only available for supervisor users. However, supervised users will be able to access to it in the near future.


Both mobile and web applications communicate through the HTTP protocol using an API developed with Node.js~\footnote{\url{https://nodejs.org/es/}} and the framework Express~\footnote{\url{https://expressjs.com/es/}}. Finally, the data is managed using a MySQL database~\footnote{\url{https://www.mysql.com/}}.

\subsubsection{User Interface}

\begin{figure*}[h!]%
    \centering
    \subfloat[]{{\includegraphics[width=.245\textwidth]{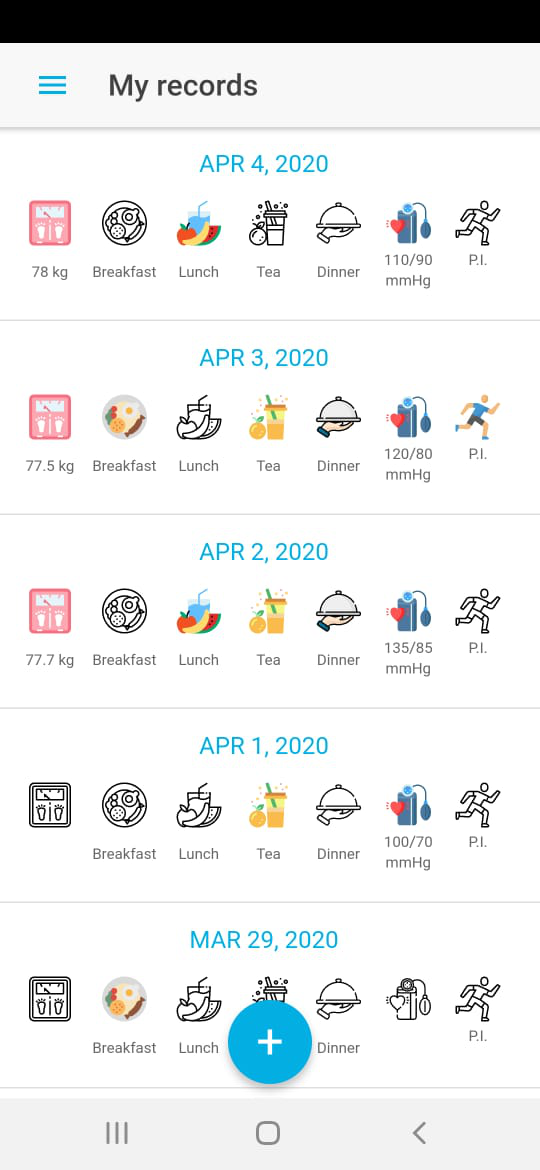} }}%
    \subfloat[]{{\includegraphics[width=.245\textwidth]{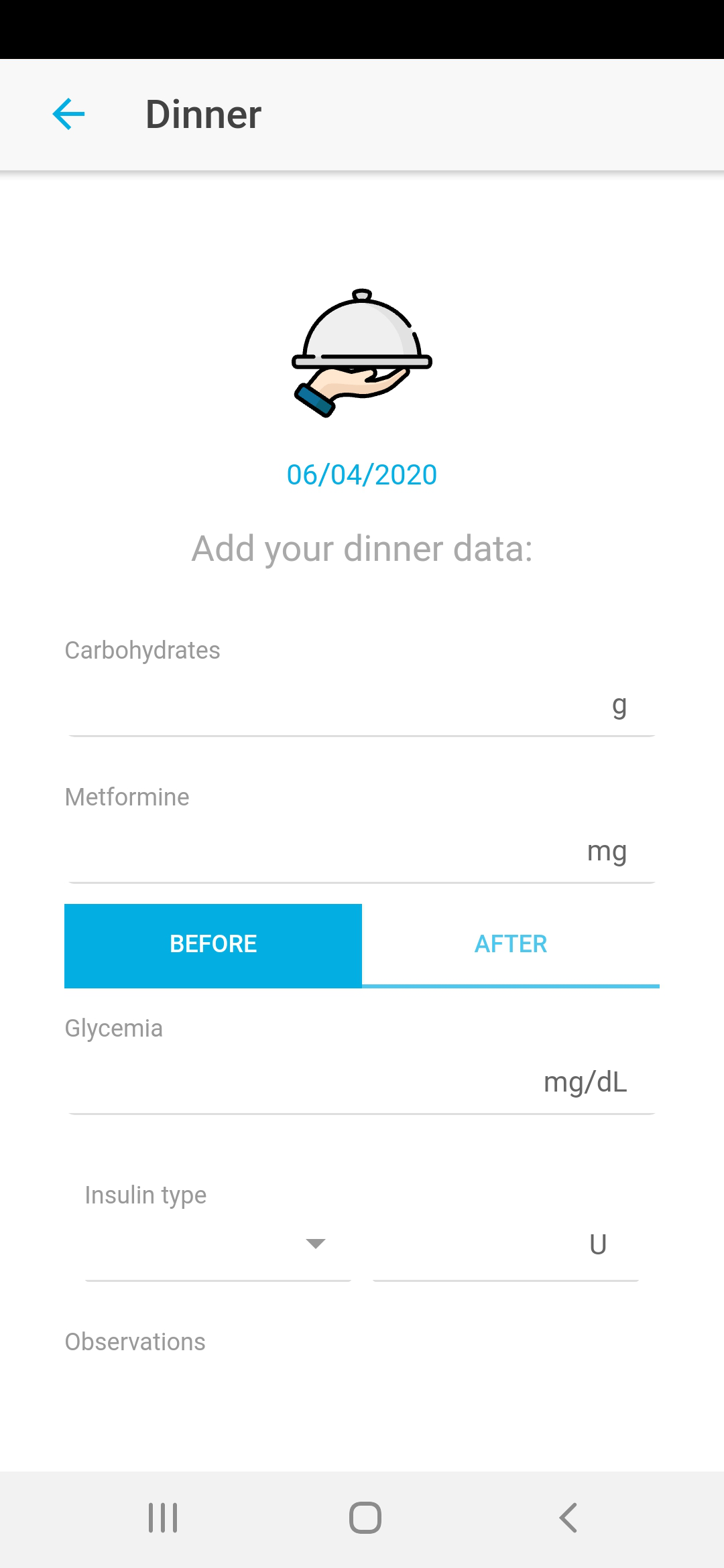} }}%
    \subfloat[]{{\includegraphics[width=.245\textwidth]{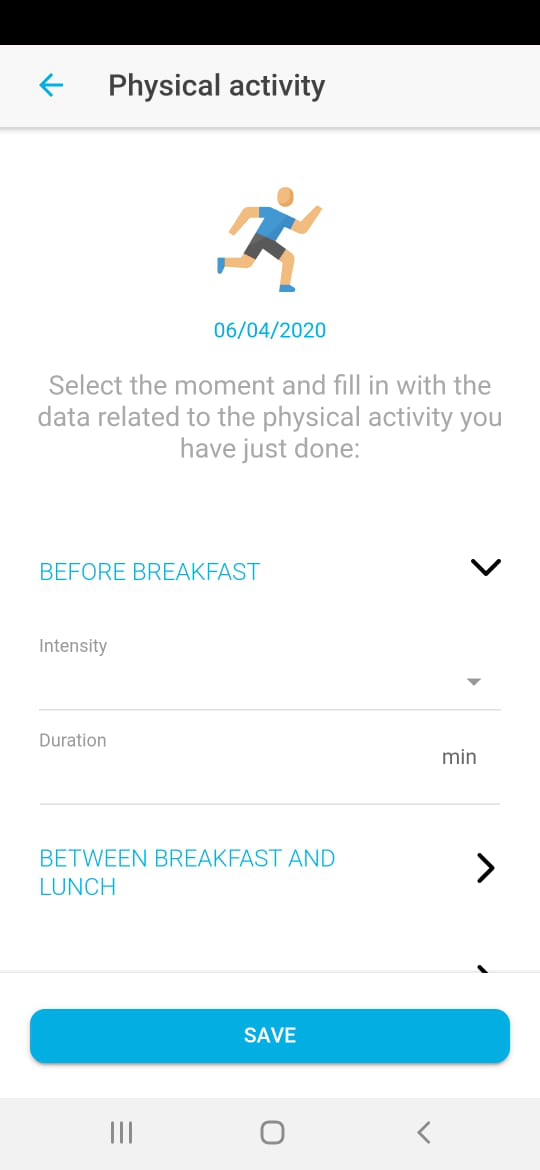}}}%
    \subfloat[]{{\includegraphics[width=.245\textwidth]{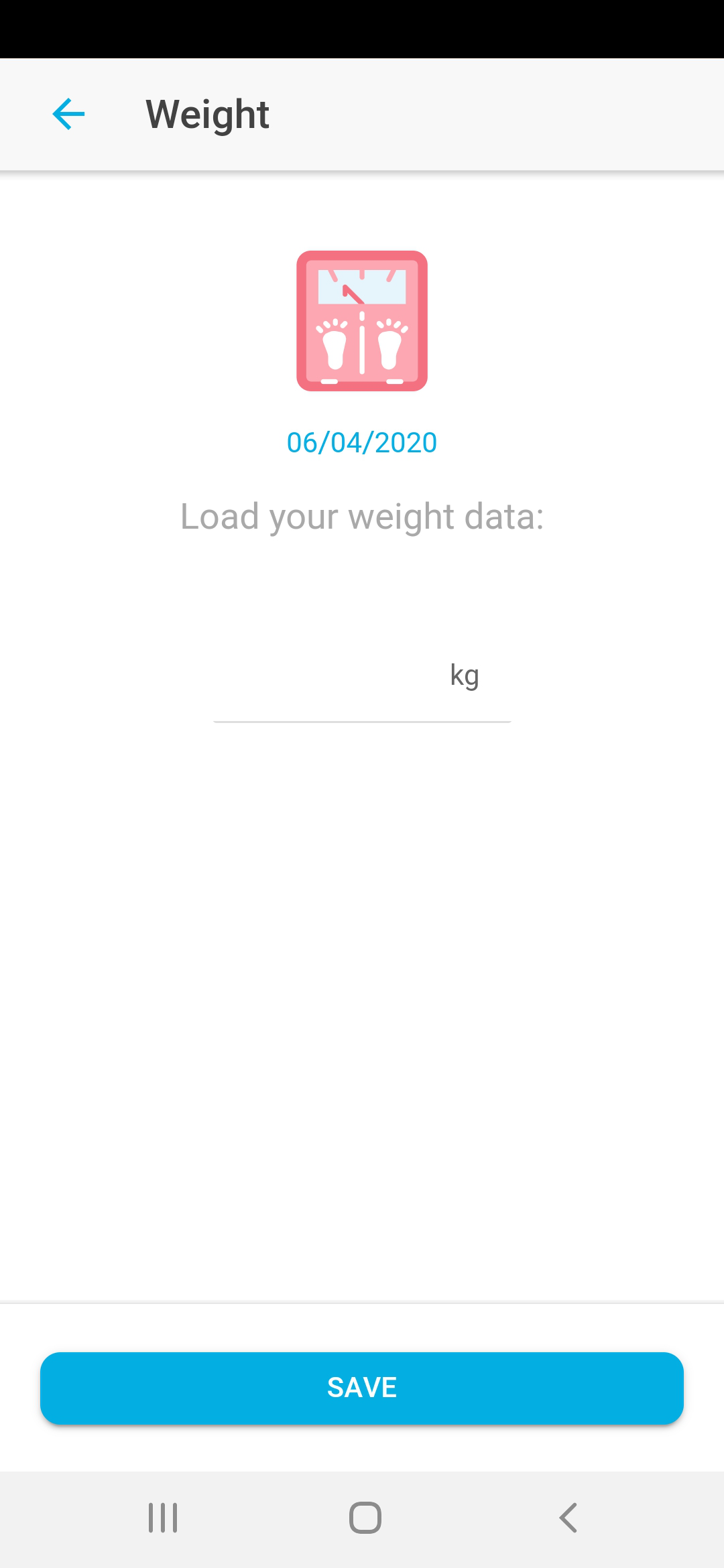} }}%
    \caption{(a-b) Implementation of FR05 and FR06. (c) Implementation of FR07. (d) Implementation of FR08. }%
    \label{fig:UI1}%
\end{figure*}

\begin{figure*}[h!]%
    \centering
    \subfloat[]{{\includegraphics[width=.245\textwidth]{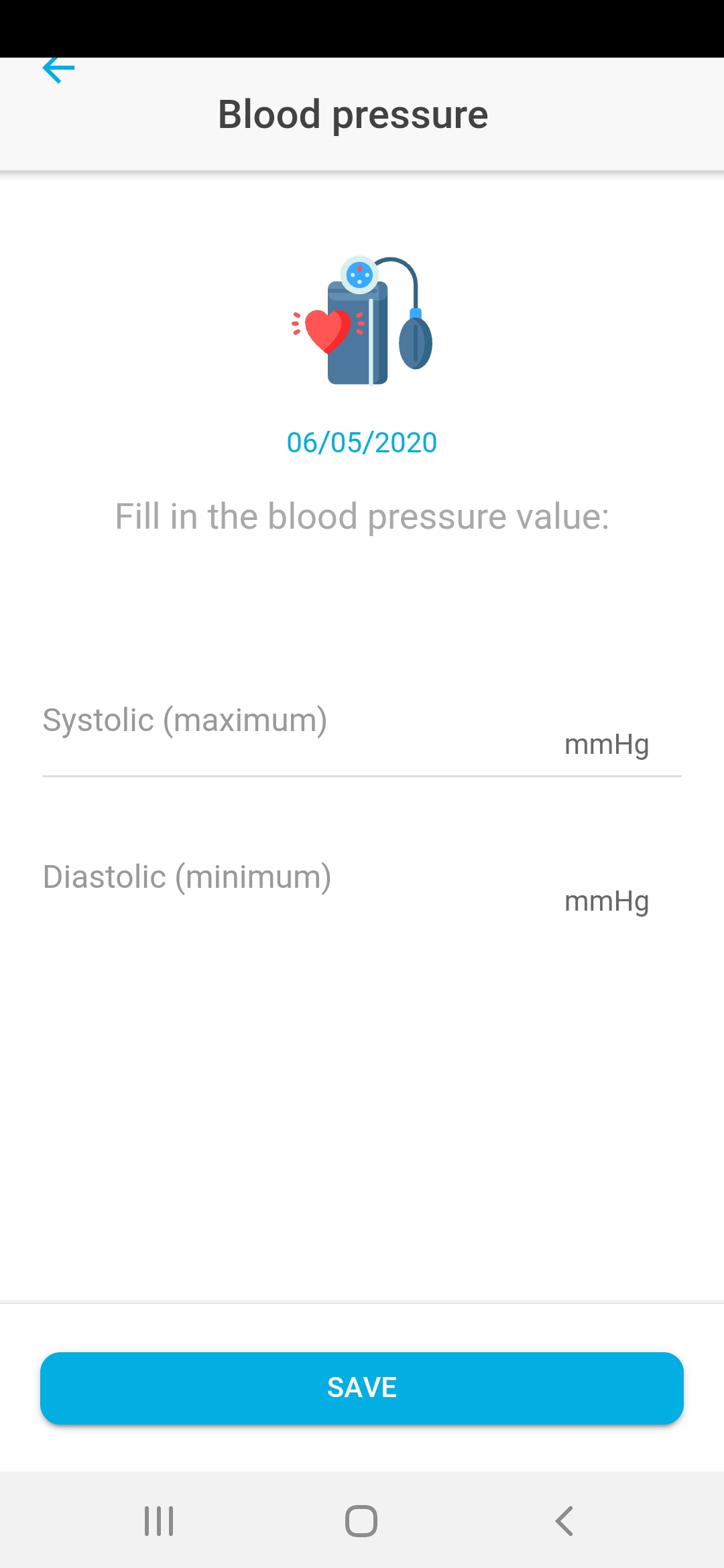} }}%
    \subfloat[]{{\includegraphics[width=.245\textwidth]{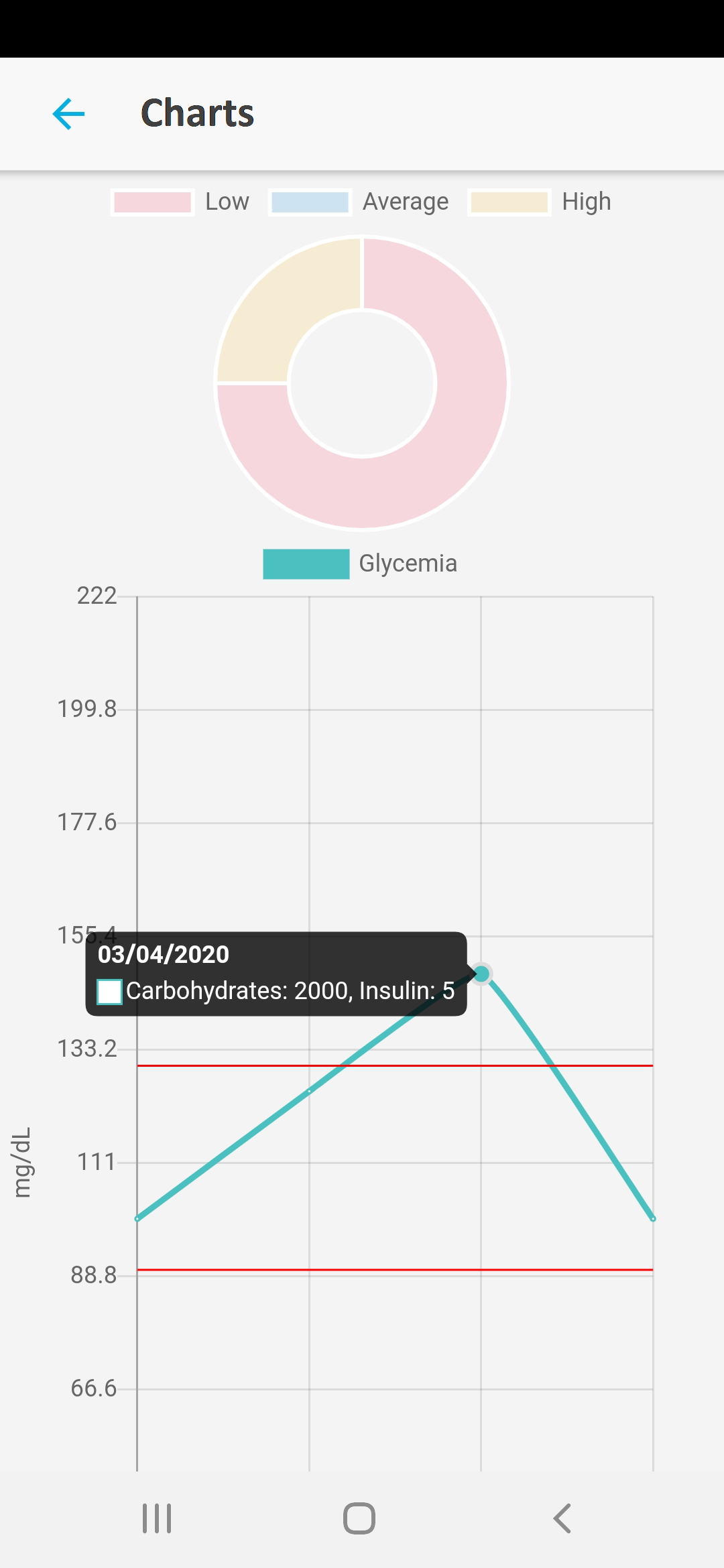} }}%
    \subfloat[]{{\includegraphics[width=.245\textwidth]{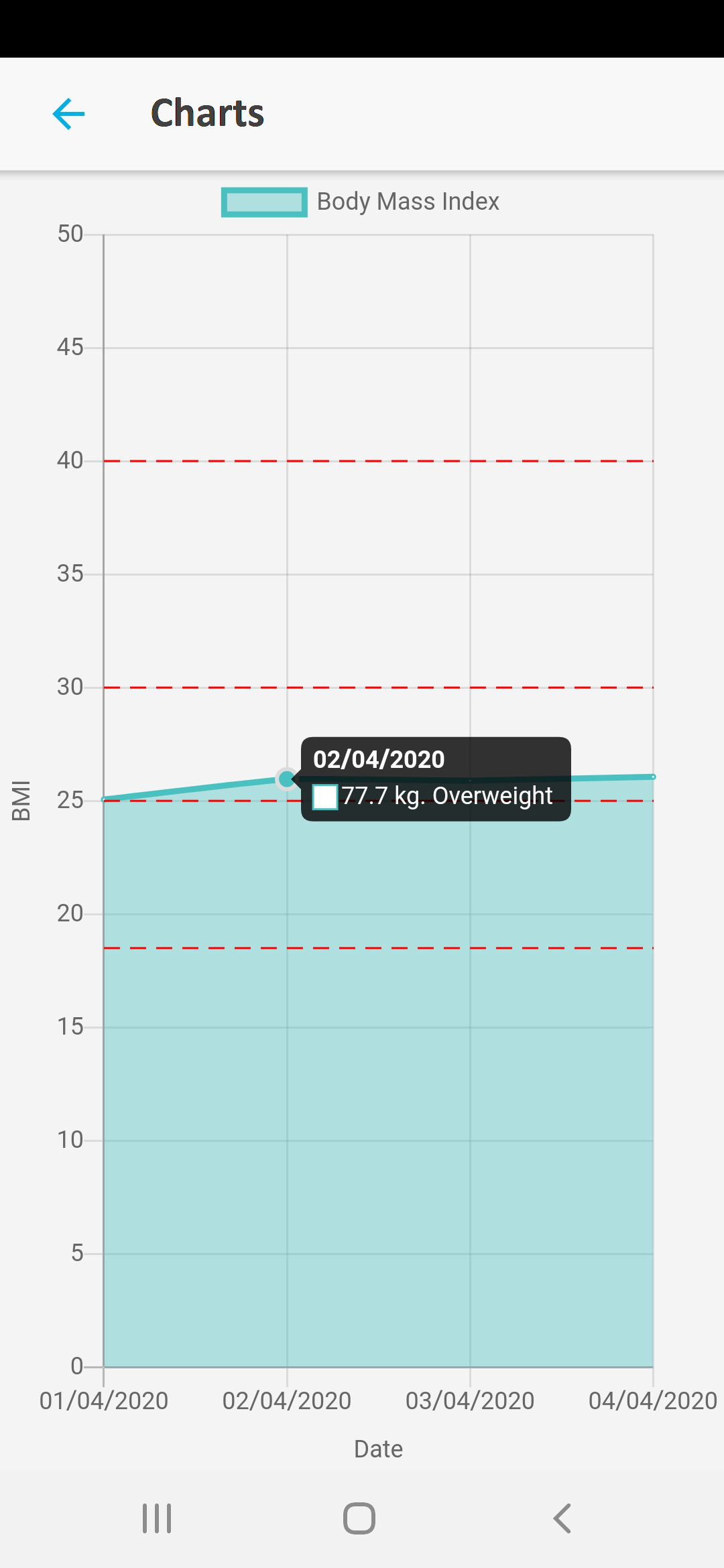} }}%
    \subfloat[]{{\includegraphics[width=.245\textwidth]{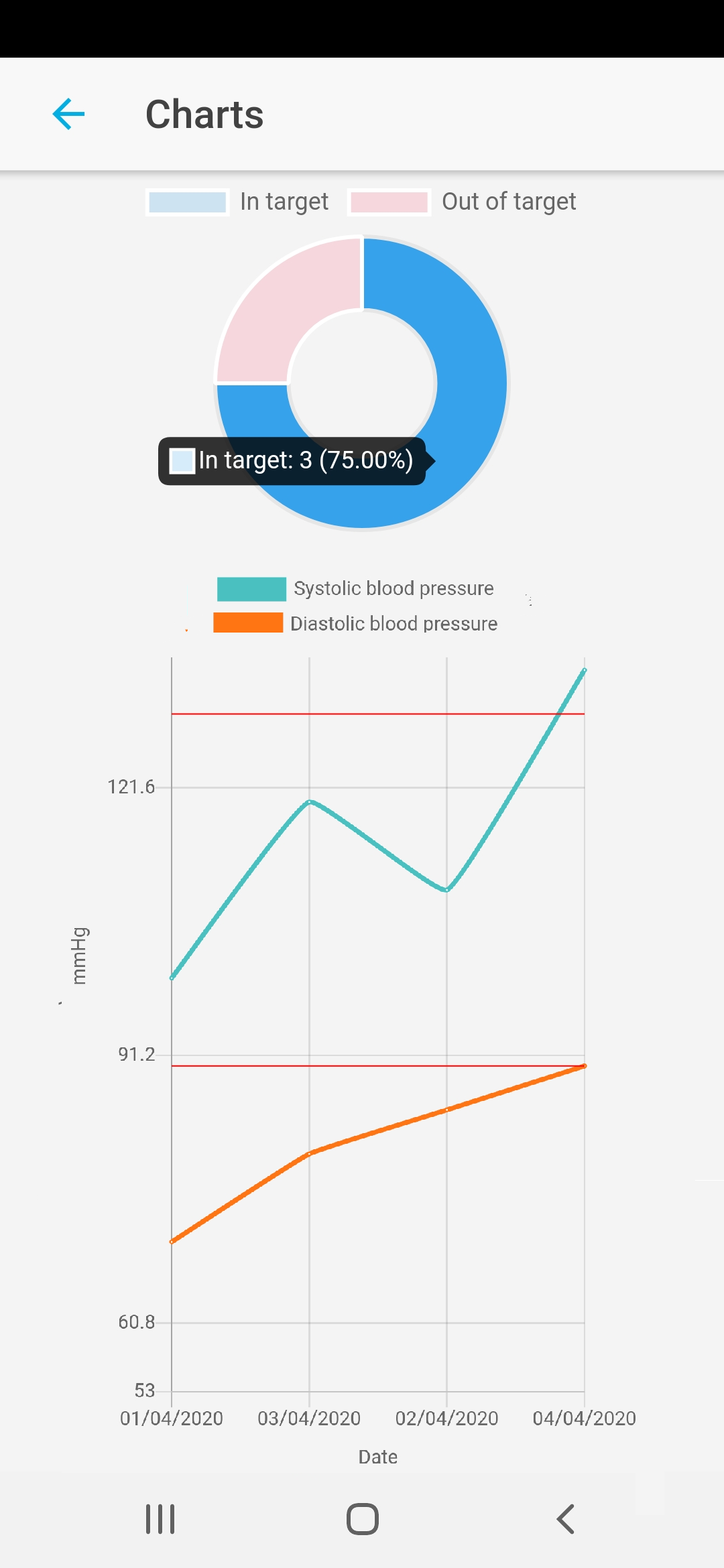} }}%
    \caption{(a) Implementation of FR09. (b) Implementation of FR10. (c) Implementation of FR11. (d) Implementation of FR12.  }%
    \label{fig:UI2}%
\end{figure*}

\begin{figure*}[h!]%
    \centering
    \subfloat[]{{\includegraphics[width=.85\textwidth]{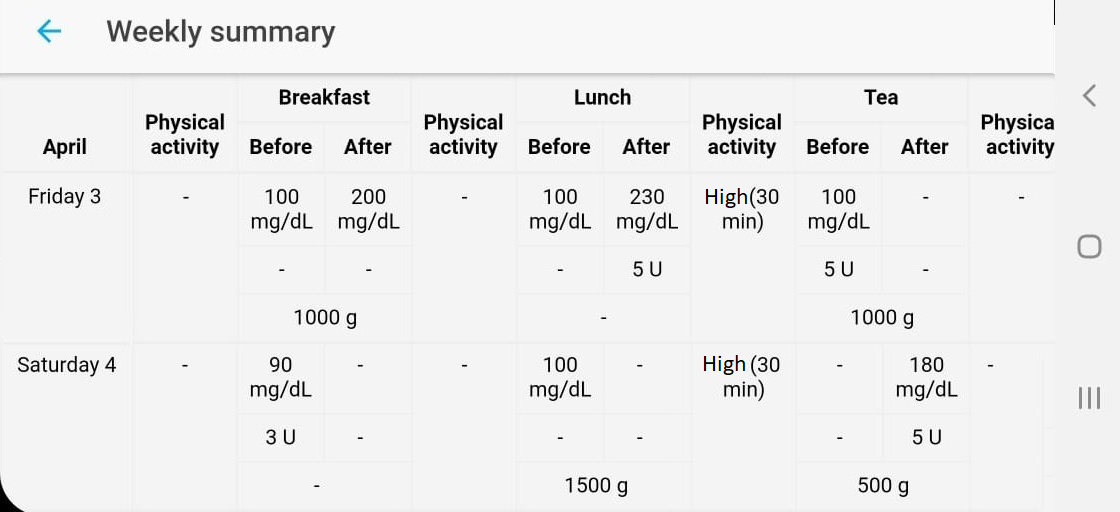} }}%
    \\
    \subfloat[]{{\includegraphics[width=.85\textwidth]{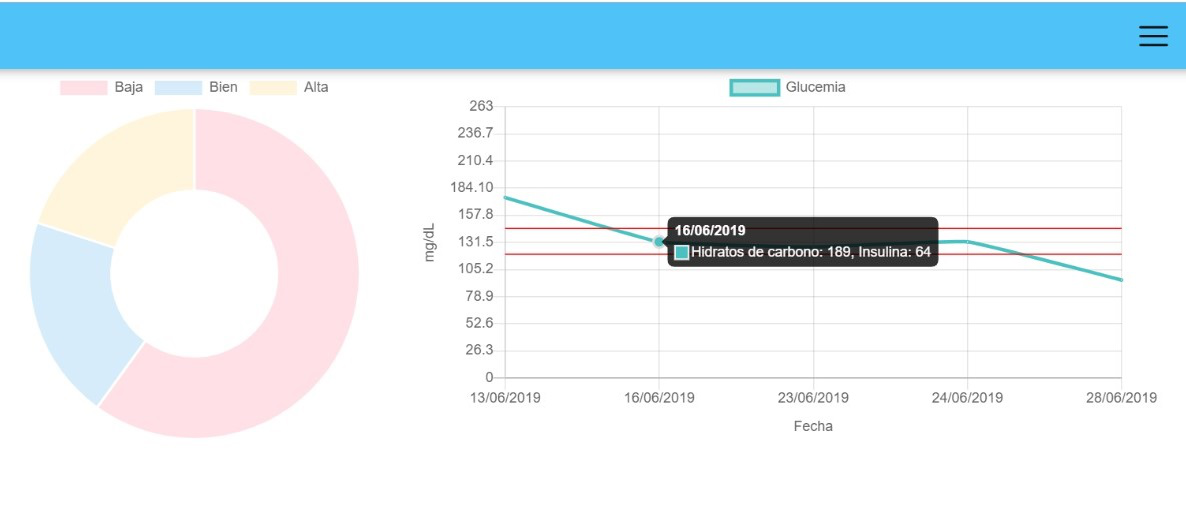} }}%

    \caption{(a) Implementation of FR13. (b) Implementation of FR10 (Web access).}%
    \label{fig:UI3}%
\end{figure*}

\begin{figure*}[h!]%
    \centering
    \subfloat[]{{\includegraphics[width=.245\textwidth]{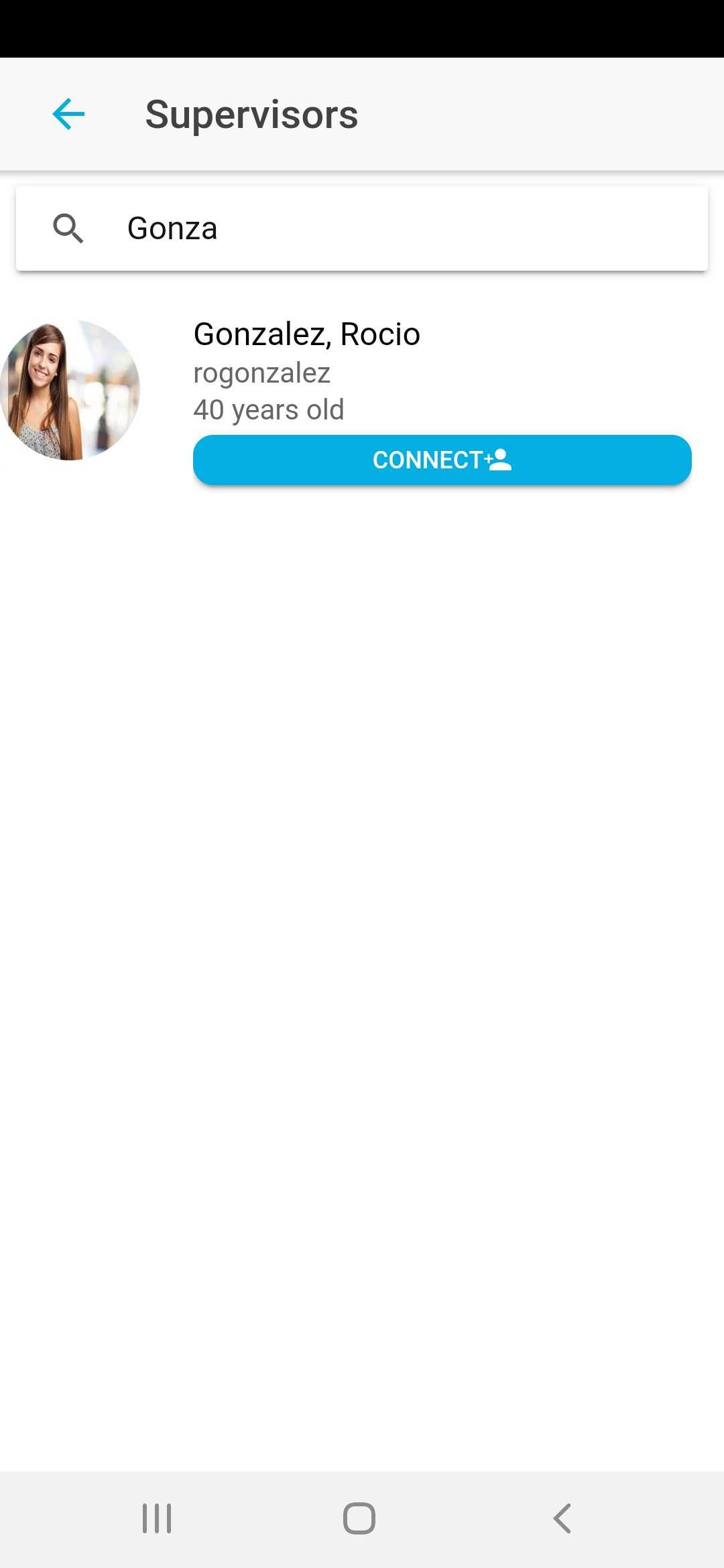} }}%
    \subfloat[]{{\includegraphics[width=.245\textwidth]{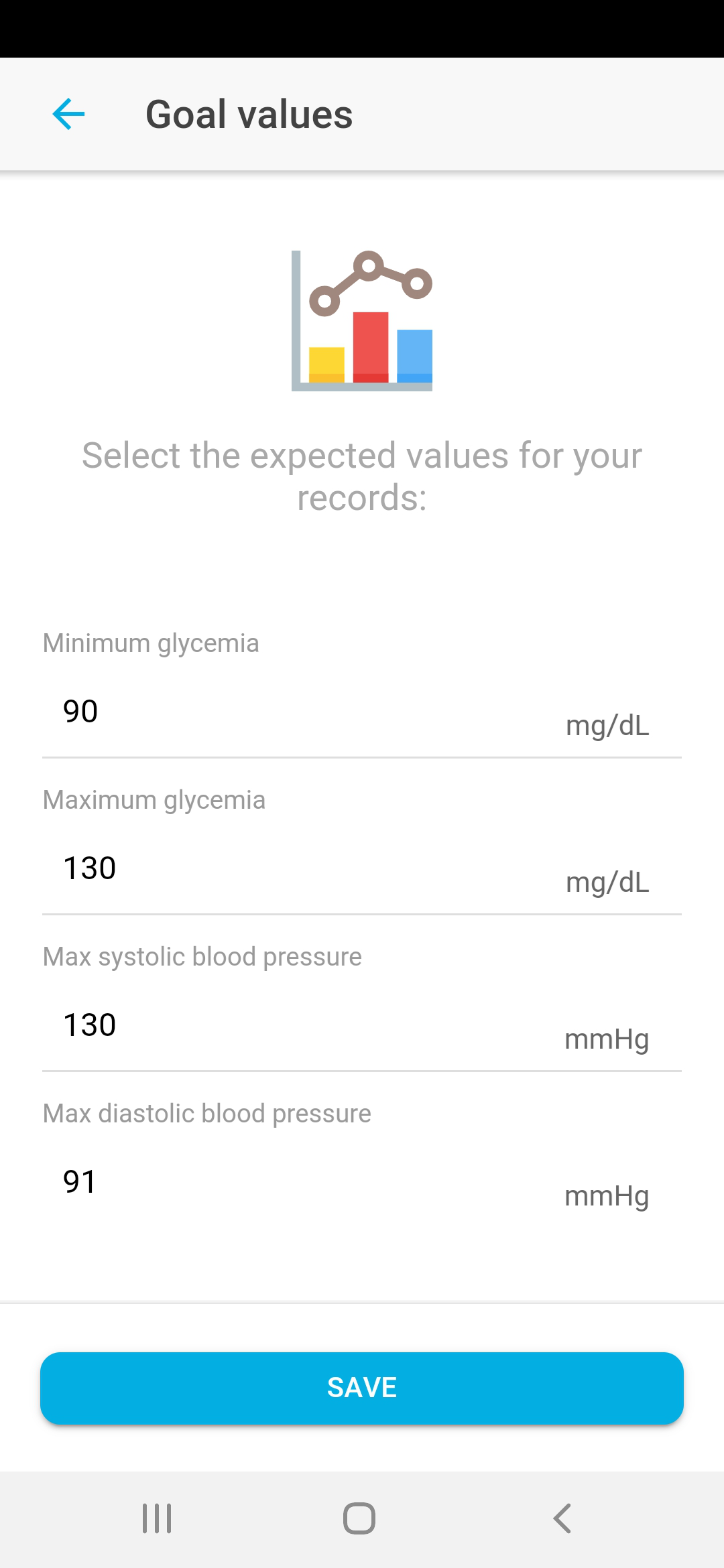} }}%
    \subfloat[]{{\includegraphics[width=.245\textwidth]{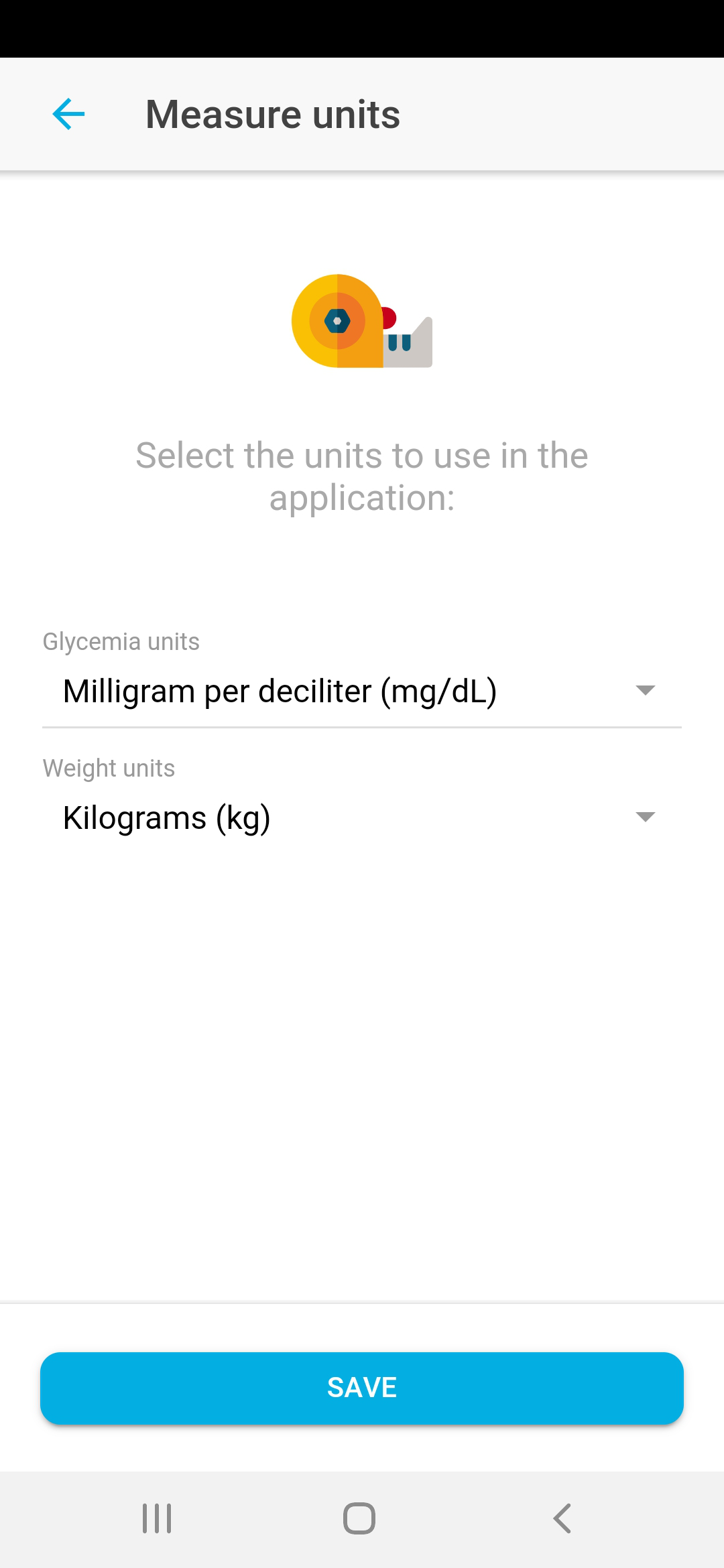} }}%
    \subfloat[]{{\includegraphics[width=.245\textwidth]{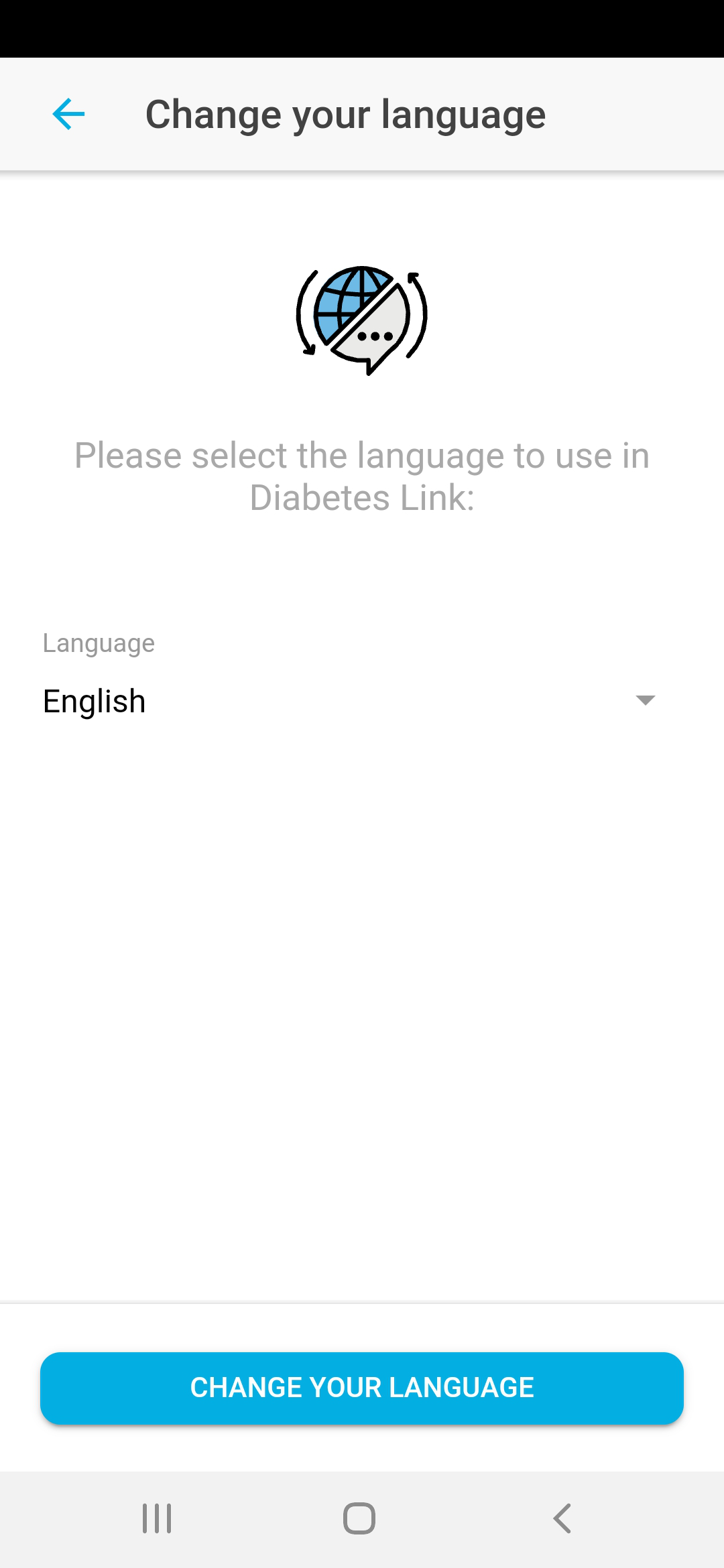} }}%
    \caption{(a) Implementation of FR15. (b) Implementation of FR19. (c) Implementation of FR20. (d) Implementation of FR21. }%
    \label{fig:UI4}%
\end{figure*}

Fig.~\ref{fig:UI1}-\ref{fig:UI4} show the implementation of several functional requirements, that were described in Table~\ref{tab:func-reqs}.

\section{Comparative Analysis of Mobile Applications for Diabetes Monitoring}
\label{sec:results}

In this section, the criteria used to survey mobile applications for DM are described, and the results found are analyzed.

\subsection{Search and Selection Criteria}

The application search and selection process for our comparative analysis was carried out following guidelines proposed in similar  studies~\cite{DiabAppSurvey3,DiabAppSurvey1,DiabAppSurvey2}. The study focused just on the Android operating system due to two reasons: first, Android is currently the most popular option, having more than 70\% of the market share~\cite{MobileOSMarketShare};  second, even though the mobile application of Diabetes Link will be soon available in iOS, now it is just present in Android.

Today, Google Play does not offer filters to apply to search results. In addition, search terms are matched against app title and/or description, which usually leads to false hits, sometimes caused by spam techniques. To ensure that our analysis is representative, the search on Google Play was carried out on a specific date (Feb. 18, 2020), and all apps found were recorded. The phrase ``diabetes control" was used since it is sufficiently general to ensure that every relevant app is detected. As a result, 246 applications were listed (incognito mode on). To be included in the analysis, an application must meet the following characteristics:

\begin{itemize}
    \item It must offer functionality that facilitates self-monitoring and control by people with diabetes;
    \item It must allow blood glucose registration as a minimum requirement;
    \item It must offer Spanish or English language interfaces.
\end{itemize}

After filtering out those apps that did not meet these requirements, 75 applications remained~\footnote{ Considering the applications that were not used, 2\% offer functionality for control, but do not support Spanish or English; 67\% are about DM, but only for educational purposes; and 31\% are not relevant to this study (poorly classified).}. In order to evaluate these applications, the first 10 were selected according to three criteria: (1) greater popularity (order of appearance on results list); (2) higher score; and (3) greater number of downloads. The 3 partial listings were combined to obtain a single final one. In total, 22 applications were reviewed (30\% of the total).

\subsection{Comparative Analysis}

After selecting the applications for the comparative analysis, their primary features were surveyed and analyzed. We have considered previous similar studies~\cite{DiabAppSurvey3,DiabAppSurvey1,DiabAppSurvey2} to determine the feature selection criteria. This comparison allows to quickly and easily see their strengths and weaknesses. 

Table~\ref{tab:ac-tab1} presents the main features of each app regarding Google Play information (order of appearence (\#), number of downloads, score and version), web access support, its adquisition cost, its communication features, and if it offers diabetes information. 
First of all, it can be noted that  just 32\% of the apps offers web access. As described in Section~\ref{sec:dlink}, Diabetes Link is cross-platform and both mobile and web accesses are supported. The former becomes more significant at the time of analizying  the information and statistical graphics, since its more user-friendly view.


In relation to adquisition costs, 78\% of the applications studied require paying a monthly or annual fee to access advanced features (for instance, calculating statistics based on the information entered, adjusting target values, exporting reports or sending them by e-mail, supporting connectivity, among others). As regards connectivity, just four applications (18\%) offer this star feature but only one does it for free. Other four applications do not offer any feature for communication support while the rest (14 apps) rely on generating PDF/XLS reports or sending them by e-mail. Diabetes Link, on its behalf, supports connectivity and offers all features for free.

\begin{sidewaystable}

\centering
\caption{\label{tab:ac-tab1} Google Play information and features of selected diabetes applications.}
\resizebox{\textwidth}{!}{%
\begin{tabular}{>{\centering\arraybackslash}p{0.24\textwidth}|>{\centering\arraybackslash}p{0.05\textwidth}>{\centering\arraybackslash}p{0.09\textwidth}>{\centering\arraybackslash}p{0.05\textwidth}>{\centering\arraybackslash}p{0.065\textwidth}|>{\centering\arraybackslash}p{0.07\textwidth}|>{\centering\arraybackslash}p{0.07\textwidth}|>{\centering\arraybackslash}p{0.085\textwidth}>{\centering\arraybackslash}p{0.15\textwidth}|>{\centering\arraybackslash}p{0.08\textwidth}}

\multicolumn{1}{c}{} & \multicolumn{4}{c}{\textbf{Google Play}} & \textbf{Web} &    & \multicolumn{2}{c}{\textbf{Communication}} & \textbf{Diabetes}  \\
\multicolumn{1}{c}{\multirow{-2}{*}{\textbf{Application}}} & \textbf{\#} & \textbf{Downloads} & \textbf{Score} & \textbf{Version} &  \textbf{Access} &  \multirow{-2}{*}{\textbf{Cost}} & \textbf{Connectivity} & \textbf{E-mail / External report} & \textbf{Info} \\ \hline
 \href{https://play.google.com/store/apps/details?id=com.mysugr.android.companion}{mySugr} & 5 & 100K & 4.6 & 3.66.1 & & PVIP~\footnote{Premium version is paid} & &\checkmark& \\
\href{https://play.google.com/store/apps/details?id=com.lifescan.reveal}{OneTouch Reveal} & 63 & 100K & 4.1 & 4.5 & & Free & & &\checkmark\\
\href{https://play.google.com/store/apps/details?id=eu.smartpatient.mytherapy}{Alarma de Medicina} & 118 & 1M & 4.7 & 3.61.1 &\checkmark(Beta) & Free &\checkmark(Beta) &\checkmark& \\
\href{https://play.google.com/store/apps/details?id=com.szyk.diabetes}{Diabetes - Diario de glucosa} & 4 & 500K & 4.7 & 4.1.5 & & PVIP & &\checkmark& \\
\href{https://play.google.com/store/apps/details?id=com.mydiabetes}{Diabetes:M} & 8 & 500K & 4.4 & 8.0 &\checkmark& PVIP &\checkmark&\checkmark&\checkmark\\
\href{https://play.google.com/store/apps/details?id=today.onedrop.android}{One Drop} & 23 & 500K & 4 & 2.0.0 & & Free & &\checkmark& \\
\href{https://play.google.com/store/apps/details?id=com.creamsoft.mygi}{Índice y Carga Glucémicos} & 115 & 500K & 4.3 & 3.3.1 & & PVIP & &\checkmark& \\
\href{https://play.google.com/store/apps/details?id=com.gexperts.ontrack}{OnTrack Diabetes} & 125 & 500K & 3.6 & 3.2.7 & & Free & &\checkmark& \\
\href{https://play.google.com/store/apps/details?id=net.klier.blutdruck}{BloodPressureDB} & 207 & 500K & 4.4 & 6.2.7 &\checkmark& PVIP & &\checkmark& \\
\href{https://play.google.com/store/apps/details?id=com.socialdiabetes.android}{SocialDiabetes} & 1 & 100K & 4.7 & 4.15 &\checkmark& PVIP &\checkmark&\checkmark& \\
\href{https://play.google.com/store/apps/details?id=melstudio.msugar}{Açúcar no sangue} & 30 & 100K & 4.9 & 3.2.5 & & PVIP & &\checkmark& \\
\href{https://play.google.com/store/apps/details?id=mobi.littlebytes.android.bloodglucosetracker}{Blood Glucose Tracker} & 21 & 100K & 4.7 & 1.8.12 & & PVIP & &\checkmark& \\
\href{https://play.google.com/store/apps/details?id=com.wonggordon.bgmonitor}{BG Monitor Diabetes} & 73 & 50K & 4.6 & 8.0.17 & & PVIP & &\checkmark& \\
\href{https://play.google.com/store/apps/details?id=com.androidbash.androidbashfirebaseupdateds}{RT Diabetes} & 171 & 500 & 4.7 & 2.03 & &  PVIP~\footnote{Trial version is free} & &\checkmark&\checkmark\\
\href{https://play.google.com/store/apps/details?id=com.msint.bloodsugar.tracker}{Diario de diabetes} & 33 & 50K & 4.6 & 1.2 & & PVIP & &\checkmark&\checkmark\\
\href{https://play.google.com/store/apps/details?id=care.sugarfree}{SugarFree} & 34 & 100 & 4.6 & 1.2.22 & & Free & & & \\
\href{https://play.google.com/store/apps/details?id=com.lehreer.glucose}{Control de la Glucosa} & 2 & 50K & 4.5 & 1.81 & & PVIP & &\checkmark&\checkmark\\
\href{https://play.google.com/store/apps/details?id=gr.tessera.fordiabetesapp}{forDiabetes} & 7 & 10K & 4.3 & 1.15.0 &\checkmark& PVIP & &\checkmark& \\
\href{https://play.google.com/store/apps/details?id=migi.app.diabetes}{Diabetes Tracker} & 9 & 100K & 3.6 & 1.11 & & PVIP & & &\checkmark\\
\href{https://play.google.com/store/apps/details?id=com.sidiary.app}{SiDiary Diabetes Management} & 10 & 50K & 4.4 & 1.45 &\checkmark& PVIP & & & \\
\href{https://play.google.com/store/apps/details?id=com.quohealth.gluco}{gluQUO} & 12 & 100K & 4.3 & 2.8.0 & & PVIP & &\checkmark& \\
\href{https://play.google.com/store/apps/details?id=com.squaremed.diabetesconnect.android}{Diabetes Connect} & 16 & 100K & 4.4 & 2.5.0 &\checkmark& PVIP &\checkmark&\checkmark& \\
\hline
\end{tabular}%
}
\end{sidewaystable}

\begin{sidewaystable}
\centering
\caption{\label{tab:ac-tab2} Features for data register and analysis of selected diabetes applications.}

\begin{tabular}
{>{\centering\arraybackslash}p{0.23\textwidth}|>{\centering\arraybackslash}p{0.07\textwidth}>{\centering\arraybackslash}p{0.07\textwidth}>{\centering\arraybackslash}p{0.07\textwidth}>{\centering\arraybackslash}p{0.07\textwidth}>{\centering\arraybackslash}p{0.08\textwidth}>{\centering\arraybackslash}p{0.07\textwidth}>{\centering\arraybackslash}p{0.07\textwidth}|>{\centering\arraybackslash}p{0.07\textwidth}>{\centering\arraybackslash}p{0.07\textwidth}>{\centering\arraybackslash}p{0.08\textwidth}>{\centering\arraybackslash}p{0.08\textwidth}}

{ }                                                   & \multicolumn{7}{c}{{ \textbf{Data register}}}                                                                                                                                                                                                                                                                      & \multicolumn{4}{c}{{ \textbf{Data analysis}}}                                                                                                                         \\
\multirow{-2}{*}{{ \textbf{Application}}}                     & { \textbf{Blood glucose}}           & { \textbf{Insulin}}          & { \textbf{Carbo-hydrates}}         & { \textbf{Body weight}}            & { \textbf{Blood pressure}}           & { \textbf{Medi- cations}}          & { \textbf{Physical activity}}         & { \textbf{Blood glucose}}           & { \textbf{Body weight}}            & { \textbf{Blood pressure}}           & { \textbf{Tabular summary}}           \\
\hline
\href{https://play.google.com/store/apps/details?id=com.mysugr.android.companion}{mySugr} & \checkmark  & \checkmark  & \checkmark  & \checkmark  & \checkmark  & \checkmark  & \checkmark  & \checkmark  &    &    & \checkmark  \\
\href{https://play.google.com/store/apps/details?id=com.lifescan.reveal}{OneTouch Reveal}                 & \checkmark  & \checkmark  & \checkmark  &    &    &    & \checkmark  & \checkmark  &    &    & \checkmark  \\
\href{https://play.google.com/store/apps/details?id=eu.smartpatient.mytherapy}{Alarma de Medicina}  & \checkmark  & \checkmark  &    & \checkmark  & \checkmark  & \checkmark  & \checkmark  & \checkmark  & \checkmark  & \checkmark  &    \\
\href{https://play.google.com/store/apps/details?id=com.szyk.diabetes}{Diabetes - Diario de glucosa}    & \checkmark  &    &    & \checkmark  &    &    &    & \checkmark  & \checkmark  &    & \checkmark  \\
\href{https://play.google.com/store/apps/details?id=com.mydiabetes}{Diabetes:M}  & \checkmark  & \checkmark  & \checkmark  & \checkmark  & \checkmark  & \checkmark  & \checkmark  & \checkmark  & \checkmark  &    &    \\
\href{https://play.google.com/store/apps/details?id=today.onedrop.android}{One Drop}          & \checkmark  & \checkmark  & \checkmark  & \checkmark  & \checkmark  & \checkmark  & \checkmark  & \checkmark  &    &    & \checkmark  \\
\href{https://play.google.com/store/apps/details?id=com.creamsoft.mygi}{Índice y Carga Glucémicos}  & \checkmark  &    & \checkmark  & \checkmark  &    &    &    & \checkmark  & \checkmark  &    &    \\
\href{https://play.google.com/store/apps/details?id=com.gexperts.ontrack}{OnTrack Diabetes}                 & \checkmark  & \checkmark  & \checkmark  & \checkmark  & \checkmark  & \checkmark  & \checkmark  & \checkmark  & \checkmark  & \checkmark  &    \\
\href{https://play.google.com/store/apps/details?id=net.klier.blutdruck}{BloodPressureDB}                & \checkmark  &    &    & \checkmark  & \checkmark  & \checkmark  &    & \checkmark  & \checkmark  & \checkmark  &    \\
\href{https://play.google.com/store/apps/details?id=com.socialdiabetes.android}{SocialDiabetes}      & \checkmark  & \checkmark  & \checkmark  & \checkmark  & \checkmark  & \checkmark  & \checkmark  & \checkmark  & \checkmark  & \checkmark  &    \\
\href{https://play.google.com/store/apps/details?id=melstudio.msugar}{Açúcar no sangue}    & \checkmark  & \checkmark  &    & \checkmark  & \checkmark  & \checkmark  &    & \checkmark  & \checkmark  & \checkmark  &    \\
\href{https://play.google.com/store/apps/details?id=mobi.littlebytes.android.bloodglucosetracker}{Blood Glucose Tracker}          & \checkmark  & \checkmark  & \checkmark  & \checkmark  & \checkmark  & \checkmark  &    & \checkmark    &    &    &    \\
\href{https://play.google.com/store/apps/details?id=com.wonggordon.bgmonitor}{BG Monitor Diabetes}                 & \checkmark      & \checkmark  & \checkmark  &      &      & \checkmark  & \checkmark  & \checkmark  &      &      &      \\
\href{https://play.google.com/store/apps/details?id=com.androidbash.androidbashfirebaseupdateds}{RT Diabetes}  & \checkmark  & \checkmark  &    & \checkmark  & \checkmark  & \checkmark  &    & \checkmark  & \checkmark  &    &    \\
\href{https://play.google.com/store/apps/details?id=com.msint.bloodsugar.tracker}{Diario de diabetes}               & \checkmark & \checkmark &   & \checkmark & \checkmark & \checkmark &   & \checkmark & \checkmark &   & \checkmark \\
\href{https://play.google.com/store/apps/details?id=care.sugarfree}{SugarFree}                & \checkmark & \checkmark & \checkmark &   & \checkmark & \checkmark & \checkmark & \checkmark &   &   &   \\
\href{https://play.google.com/store/apps/details?id=com.lehreer.glucose}{Control de la Glucosa}                            & \checkmark & \checkmark &   &   &   & \checkmark &   & \checkmark &   &   &   \\
\href{https://play.google.com/store/apps/details?id=gr.tessera.fordiabetesapp}{forDiabetes} & \checkmark & \checkmark & \checkmark & \checkmark & \checkmark & \checkmark & \checkmark & \checkmark & \checkmark & \checkmark &   \\
\href{https://play.google.com/store/apps/details?id=migi.app.diabetes}{Diabetes Tracker}                                 & \checkmark &   &   &   & \checkmark & \checkmark & \checkmark & \checkmark &   & \checkmark &   \\
\href{https://play.google.com/store/apps/details?id=com.sidiary.app}{SiDiary Diabetes Management}                        & \checkmark & \checkmark & \checkmark & \checkmark & \checkmark & \checkmark &   & \checkmark & \checkmark & \checkmark &   \\
\href{https://play.google.com/store/apps/details?id=com.quohealth.gluco}{gluQUO}                      & \checkmark & \checkmark & \checkmark & \checkmark &   &   & \checkmark & \checkmark &   &   &   \\
\href{https://play.google.com/store/apps/details?id=com.squaremed.diabetesconnect.android}{Diabetes Connect}                                 & \checkmark & \checkmark & \checkmark & \checkmark & \checkmark & \checkmark & \checkmark & \checkmark & \checkmark & \checkmark & \checkmark \\
\hline
\end{tabular}%

\end{sidewaystable}

Table~\ref{tab:ac-tab2} summarizes the funcionality features of each application regarding data register and statistical analysis. Each app was surveyed to determine if it supports recording blood glucose, insulin, carbo-hydrates, body weight, blood pressure, medication, and physical activity. Additionally, wether the application analyzes blood glucose,  body weight, and blood pressure data or not. Last, if it offers a tabular summary of registered data or not.
 
Between 59\% and 82\% of the selected applications allow recording measurements for blood glucose, insulin, carbohydrate intake, physical activity, food and medications intake. In addition, ~77\% allow recording body weight, and less than 73\% allow recording blood pressure. As for the ability of adding labels or notes to the records, all applications offer this feature. Diabetes Link, on the other hand, allows recording of all of these measurements.

All of the applications that were reviewed offer statistical charts/tables for blood glucose measurements. As regards the other parameters, the percentages of applications that analyze them are 59\% and 41\%, they correspond to body weight and blood pressure, respectively. Furthermore, only 27\% are capable of generating summary tables to report all recorded values. In contrast, Diabetes Link generates statistics for all these parameters, as well as a weekly summary tables for blood glucose, insulin and carbohydrate intake measurements, as well as physical activity done. 
It should be noted that the idea of adding these weekly summaries was suggested by the physicians during their interviews. They stated that they usually ask their patients to provide written records (diary) with this information to analyze patient status and adjust treatment accordingly.

From Tables~\ref{tab:ac-tab1} and~\ref{tab:ac-tab2}, it can be observed that none of the applications provides information about diabetes, statistics and connection with monitors simultaneously for free. In that context, Diabetes Link sets itself apart by providing all these features at no charge.


\begin{table}[t!]
\centering
\caption{\label{tab:ac-tab3} Features for customization of selected diabetes applications.}

\begin{tabular}
{>{\centering\arraybackslash}p{0.38\textwidth}|>{\centering\arraybackslash}p{0.14\textwidth}>{\centering\arraybackslash}p{0.19\textwidth}>{\centering\arraybackslash}p{0.27\textwidth}}

{ }                                                   & \multicolumn{3}{c}{{ \textbf{Customization}}} \\
\multirow{-2}{*}{{ \textbf{Application}}}                     & { \textbf{Language}}           & { \textbf{Target valures}}          & { \textbf{Units of measurement}}           \\
\hline
\href{https://play.google.com/store/apps/details?id=com.mysugr.android.companion}{mySugr} & \checkmark  & \checkmark  & \checkmark  \\
\href{https://play.google.com/store/apps/details?id=com.lifescan.reveal}{OneTouch Reveal} & \checkmark  & \checkmark  &   \\
\href{https://play.google.com/store/apps/details?id=eu.smartpatient.mytherapy}{Alarma de Medicina}  & \checkmark  & \checkmark  & \checkmark  \\
\href{https://play.google.com/store/apps/details?id=com.szyk.diabetes}{Diabetes - Diario de glucosa}   & \checkmark  & \checkmark  & \checkmark  \\
\href{https://play.google.com/store/apps/details?id=com.mydiabetes}{Diabetes:M}  & \checkmark  & \checkmark  & \checkmark  \\
\href{https://play.google.com/store/apps/details?id=today.onedrop.android}{One Drop}          & \checkmark  & \checkmark  & \checkmark   \\
\href{https://play.google.com/store/apps/details?id=com.creamsoft.mygi}{Índice y Carga Glucémicos}  & \checkmark  & \checkmark  & \checkmark  \\
\href{https://play.google.com/store/apps/details?id=com.gexperts.ontrack}{OnTrack Diabetes}  &   & \checkmark  & \checkmark\\
\href{https://play.google.com/store/apps/details?id=net.klier.blutdruck}{BloodPressureDB}  & \checkmark  &   & \checkmark  \\
\href{https://play.google.com/store/apps/details?id=com.socialdiabetes.android}{SocialDiabetes}  & \checkmark  & \checkmark  & \checkmark  \\
\href{https://play.google.com/store/apps/details?id=melstudio.msugar}{Açúcar no sangue}  & \checkmark  & \checkmark  & \checkmark  \\
\href{https://play.google.com/store/apps/details?id=mobi.littlebytes.android.bloodglucosetracker}{Blood Glucose Tracker}          &   & \checkmark  & \checkmark  \\
\href{https://play.google.com/store/apps/details?id=com.wonggordon.bgmonitor}{BG Monitor Diabetes}                 & \checkmark      & \checkmark  & \checkmark  \\
\href{https://play.google.com/store/apps/details?id=com.androidbash.androidbashfirebaseupdateds}{RT Diabetes}  & \checkmark  &   &       \\
\href{https://play.google.com/store/apps/details?id=com.msint.bloodsugar.tracker}{Diario de diabetes}               &  & \checkmark &  \checkmark  \\
\href{https://play.google.com/store/apps/details?id=care.sugarfree}{SugarFree}                & \checkmark & \checkmark & \checkmark  \\
\href{https://play.google.com/store/apps/details?id=com.lehreer.glucose}{Control de la Glucosa}                            &  &  &     \\
\href{https://play.google.com/store/apps/details?id=gr.tessera.fordiabetesapp}{forDiabetes} & \checkmark & \checkmark & \checkmark  \\
\href{https://play.google.com/store/apps/details?id=migi.app.diabetes}{Diabetes Tracker}                                 &  & \checkmark  &  \checkmark    \\
\href{https://play.google.com/store/apps/details?id=com.sidiary.app}{SiDiary Diabetes Management}                        & \checkmark & \checkmark & \checkmark  \\
\href{https://play.google.com/store/apps/details?id=com.quohealth.gluco}{gluQUO}                      & \checkmark & \checkmark & \checkmark   \\
\href{https://play.google.com/store/apps/details?id=com.squaremed.diabetesconnect.android}{Diabetes Connect}                                 & \checkmark & \checkmark & \checkmark  \\
\hline
\end{tabular}%

\end{table}

Finally, Table~\ref{tab:ac-tab3} shows the customization features of each application considering user language, target values and units of measurement. First, 22\% of the applications does not include multi-language support. Having this feature allows Diabetes Link to offer an expanded reach, which means more people can use it. Both target values and units of measurement can be configured in most applications; in particular, 86\% for the former and 82\% for the latter. Diabetes Link support these two configuration options allowing users to customize their level of control and preferred units of measurement.

\subsection{Limitations}



Due to the great number of existing diabetes apps and the large amount of work required to revise them, some limitation of the comparative analysis were imposed. First, the review study just considered 
the Android OS. Although Android has more than 70\% of the market share~\cite{MobileOSMarketShare}, currently available mobile applications for other OS's (especially iOS) were not considered.

In addition, the comparative analysis carried out only gathered the presence or absence of a given feature, as a way of simplifying the large collection task. In that sense, neither the quality of functions nor their effectiveness were evaluated. However, useful information could still be extracted.



\section{Conclusions and Future Work}
\label{sec:conclusions}

Adequate diabetes control is fundamental to improve the quality of life of people who suffer from this disease. This paper presented Diabetes Link, a comprehensive platform to simplify the control and monitoring of people with DM. The exhaustive comparative analysis carried out shows that Diabetes Link presents distinctive and superior features against other current proposals. In particular, we can highlight it is cross-platform, it supports connectivity, and it allows users to record/analyze various parameters relevant for the appropiate treatment  (all for free). It is for that reason that Diabetes Link is expected to help make day-to-day control easier and optimize the efficacy in DM control and treatment.

Future work focuses on the extension of the comparative analysis to include two aspects: first, considering advanced features like IoT integration; second, considering  mobile applications from iOS, especially when the iOS version of Diabetes Link is available.

%
%
%
\bibliographystyle{splncs04}
\bibliography{references}
\end{document}